\documentclass[twocolumn,twoside,slac]{revtex4}
\usepackage{graphicx}
\usepackage{fancyhdr}
\usepackage[figuresright]{rotating}
\begin{document}

\title{The GRID and the Linux Farm at the RCF}

%

\author{A. Chan, R. Hogue, C. Hollowell, O. Rind, J. Smith, T. Throwe, T. Wlodek, D. Yu}
\affiliation{Brookhaven National Laboratory, NY 11973, USA}

\begin{abstract}
The emergence of the GRID architecture and related tools will
have a large impact in the operation and design of present
and future large clusters. We present here the ongoing efforts 
to equip the Linux Farm at the RHIC Computing Facility with
Grid-like capabilities.
\end{abstract}

\maketitle

\thispagestyle{fancy}

\section{BACKGROUND}

The RHIC Computing Facility (RCF) is a large scale data processing 
facility at Brookhaven National Laboratory (BNL) for the Relativistic 
Heavy Ion Collider (RHIC), a collider dedicated to high-energy nuclear 
physics experiments. 

RHIC's first physics collisions occurred in the Summer of 2000, when 
all four experiments began recording data from the collisions. Year 3 
of RHIC operations is currently underway. 

The RCF provides for the computational needs of the five RHIC 
experiments (BRAHMS, PHENIX, PHOBOS, PP2PP and STAR), including batch, 
mail, printing and data storage. In addition, BNL is the U.S. Tier 1 
Center for ATLAS computing, and the RCF also provides for the computational 
needs of the U.S. collaborators in ATLAS. 

The Linux Farm at the RCF provides the majority of the CPU power 
in the RCF. It is currently listed as the $3^{rd}$ largest cluster, 
according to "Clusters Top500" (http://clusters.top500.org). Figure 1 
shows the rapid growth of the Linux Farm in the last few years. 

\begin{figure*}[t]
\begin{center}
\includegraphics[width=135mm]{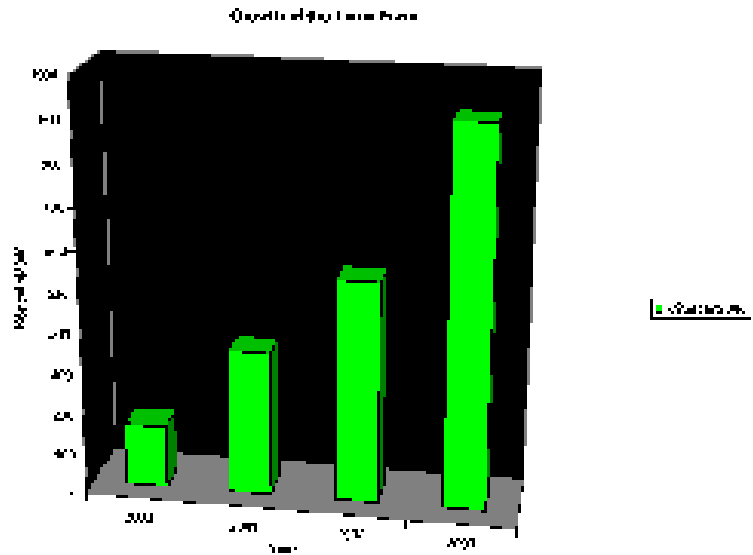}
\end{center}
\caption{The growth of the Linux Farm at the RCF.}
\end{figure*}

All aspects of its development (hardware and software), operations
and maintenance are overseen by the Linux Farm group, currently a 
staff of 5 FTE within the RCF. 
 
\section{HARDWARE}

The Linux Farm is built with commercially available thin rack-mounted,
Intel-based servers (1-U and 2-U form factors). Currently, there are 
1097 dual-CPU production servers with approximately 917,728 SpecInt2000.
Table 1 summarizes the hardware currently in service in the Linux Farm.
Hardware reliability has not been an issue at the RCF. The average 
failure rate is 0.0052 $failures/(machine \cdot month)$, which translates
to 5.7 hardware failures per month at its present size. Hardware failures
are dominated by disk and power supply failures. A detailed breakdown of
the hardware failures by category is shown in Figure 2. 

\begin{table}[t]
\begin{center}
\caption{Linux Farm hardware}
\begin{tabular}{|c|c|c|c|c|}
\hline \textbf{Brand} & \textbf{CPU} & \textbf{RAM} &
\textbf{Storage} & \textbf{Quantity}
\\
\hline VA Linux & 450 MHz & 0.5-1 GB & 9-120 GB & 154\\
\hline VA Linux & 700 MHz & 0.5 GB & 9-36 GB & 48\\
\hline VA Linux & 800 MHz & 0.5-1 GB & 18-480 GB & 168\\
\hline IBM & 1.0 GHz & 0.5-1 GB & 18-144 GB & 315\\
\hline IBM & 1.4 GHz & 1 GB & 36-144 GB & 160\\
\hline IBM & 2.4 GHz & 1 GB & 240 GB & 252\\
 \hline
\end{tabular}
\end{center}
\end{table}

\begin{figure*}[t]
\begin{center}
\includegraphics[width=135mm]{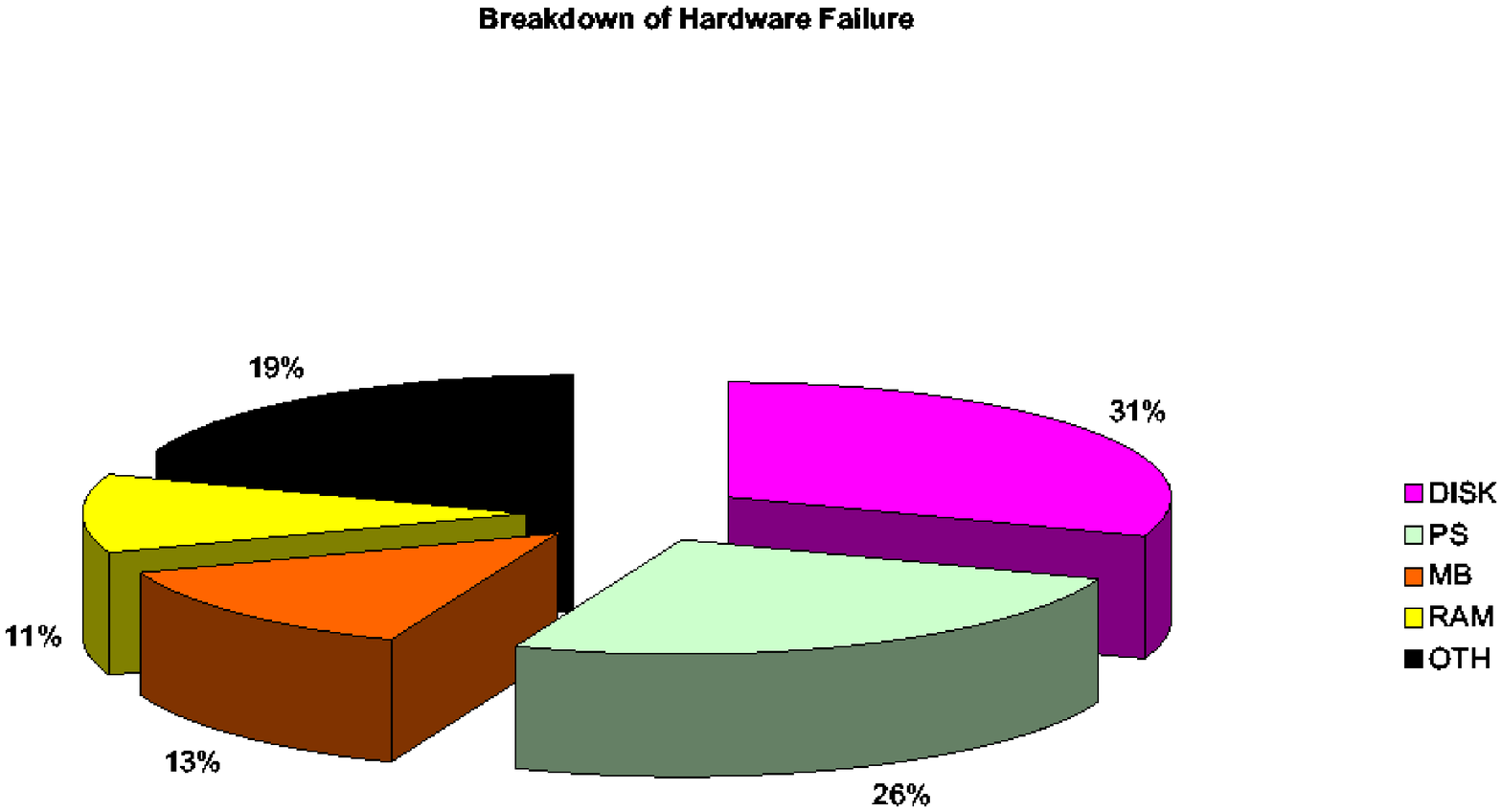}
\end{center}
\caption{Breakdown of hardware failure by category.}
\end{figure*}

\section{SOFTWARE}

The Linux Farm at the RCF uses a custom image of RedHat 7.2,
modified to conform to the requirements of the RHIC experiments 
and to the security protocols of BNL. The customized image is
installed via KickStart [1], the RedHat Linux automated installation
tool.

The Linux Farm servers are equipped with a variety of compilers
(gcc, PGI, Intel) and debuggers (gdb, Totalview, Intel) to provide
a large degree of flexibility to its end users. In addition, the
servers also support network file systems (AFS, NFS) and 
batch services, LSF and a RCF-designed software compatible with
our MDS system. Figure 3 shows the GUI for the RCF-designed
batch software. 

\begin{figure}[h]
\begin{center}
\includegraphics[width=70mm]{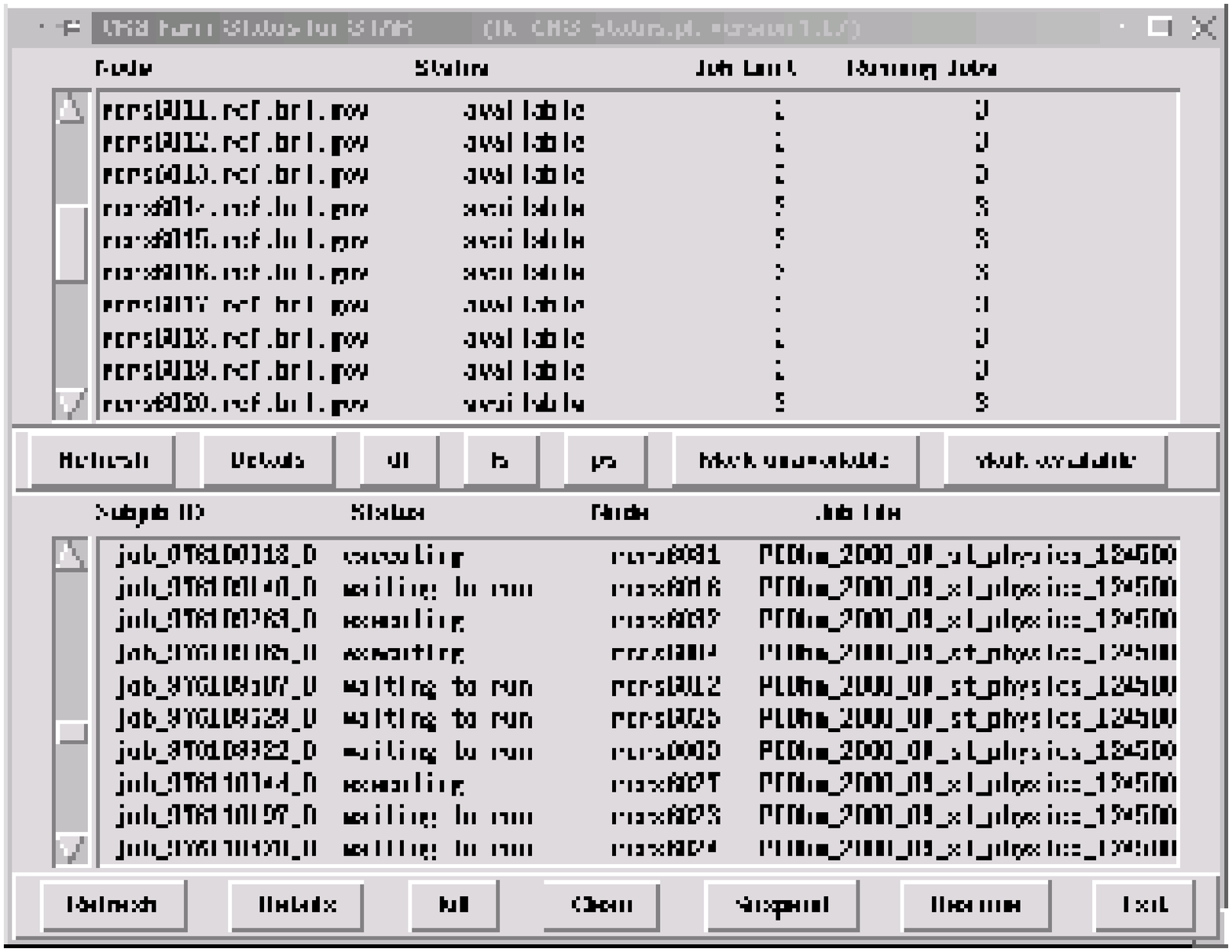}
\caption{GUI for RCF-designed batch software.}
\end{center}
\end{figure}

Monitoring and control of the cluster hardware, software and 
infrastructure (power and cooling) is provided via a mix of 
open-source software, RCF-designed software and vendor-provided
software. Figures 4 and 5 display some of our cluster monitoring 
tools, while figure 6 shows the historical temperature data of 
the cluster equipment, one of the tools we use to monitor the 
infrastructure supporting the Linux Farm. 

\begin{figure}[h]
\begin{center}
\includegraphics[width=70mm]{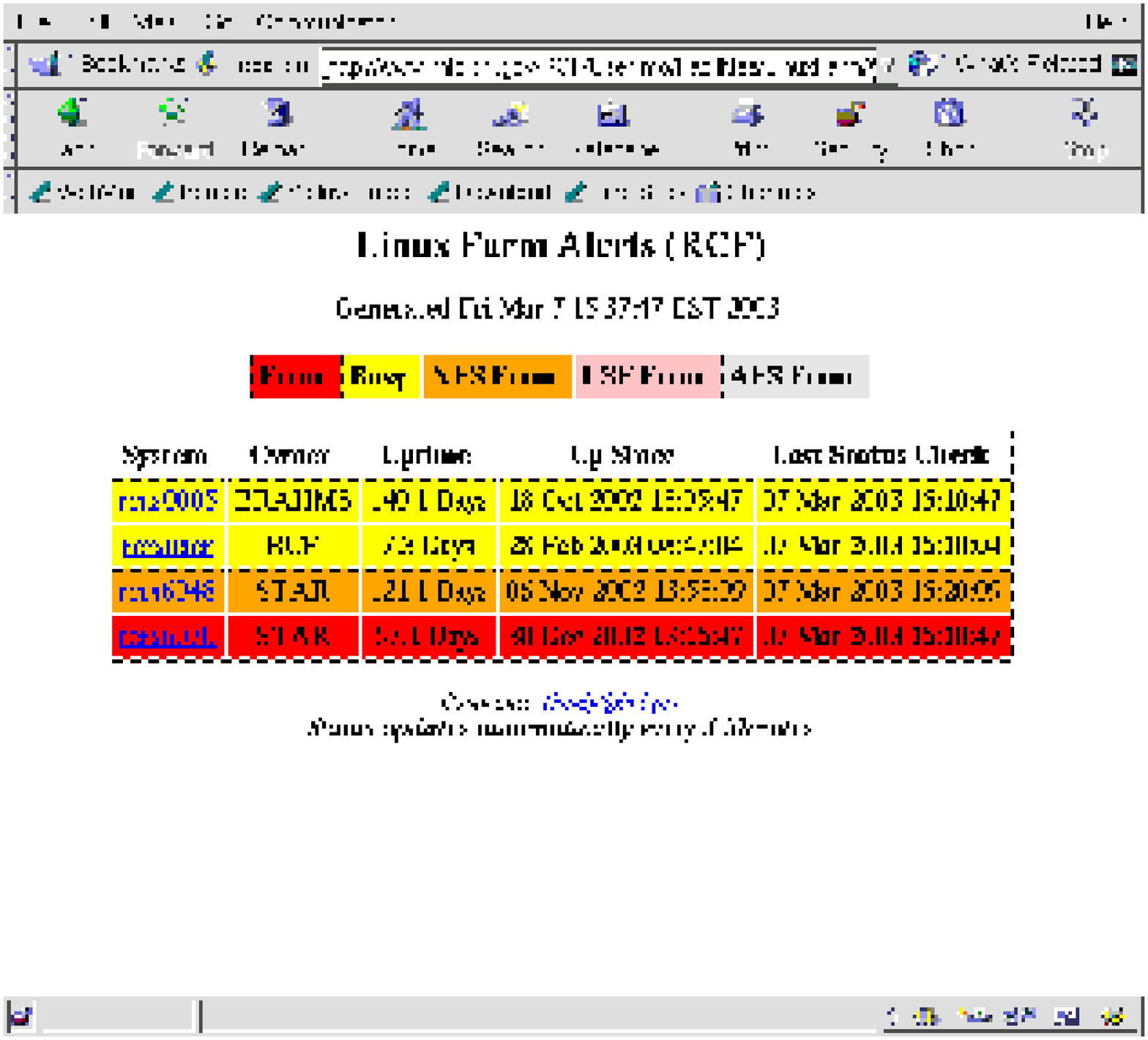}
\caption{Alert status of Linux Farm nodes.}
\end{center}
\end{figure}

\begin{figure}[h]
\begin{center}
\begin{turn}{-90}
\includegraphics[width=55mm]{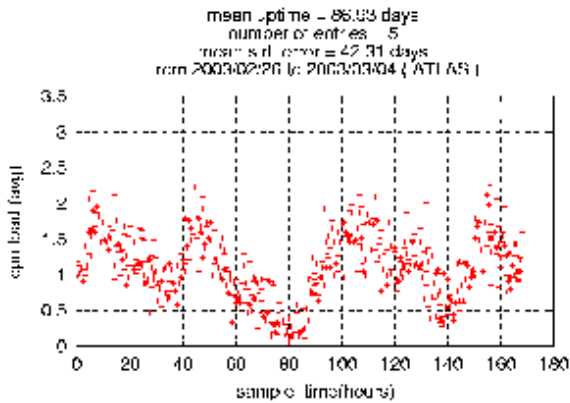}
\end{turn}
\caption{Historical load information on the ATLAS cluster.}
\end{center}
\end{figure}

\begin{figure}[h]
\begin{center}
\begin{turn}{-90}
\includegraphics[width=55mm]{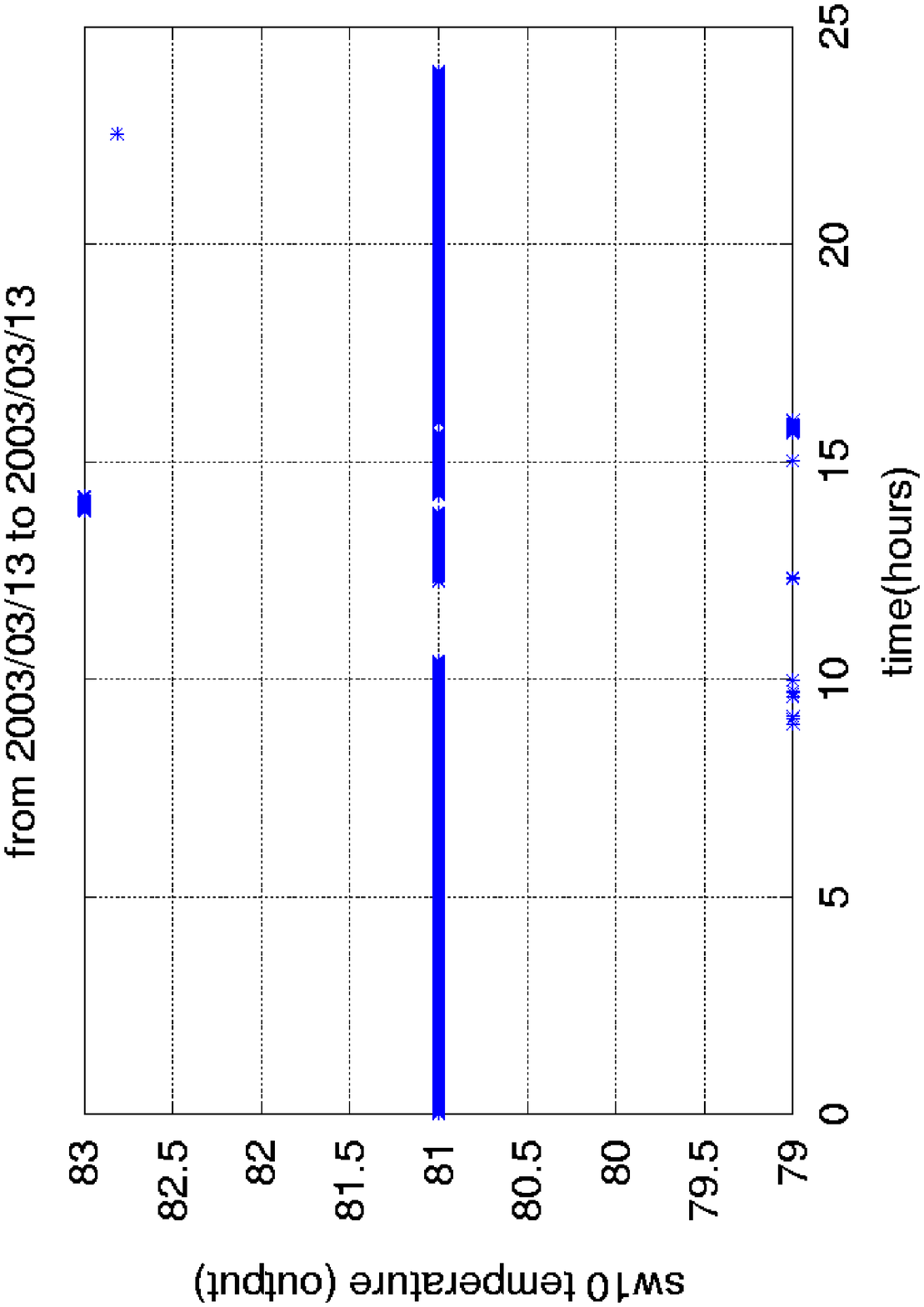}
\end{turn}
\caption{Historical temperature data in the vicinity of the Linux Farm equipment.}
\end{center}
\end{figure}

\section{SECURITY}

Because of general cyber-security standards at BNL and as part of 
the general security policy of the RCF, several measures were taken
to insure that only authorized users can access the Linux Farm.

A firewall has been placed around the RCF (including the Linux Farm) 
to minimize security breaches. In addition, users are only allowed 
interactive access to the Linux Farm via dedicated gatekeeper systems 
whose software is kept up-to-date to minimize the exploitation of
security weaknesses in the software. No other method is provided for 
interactive access to the Linux Farm. The operating system in the Linux 
Farm servers has been modified to accomodate these security measures
and to enhance our ability to detect unauthorized access. 

The RCF has also begun the deployment of a Kerberos 5 single sign-on
system that will eventually replace our current authentication system.

\section{GRID-LIKE CAPABILITIES}

GRID-like technology has evolved from conceptual designs to promising
prototypes with real capabilities in the last few years. It is a natural
fit to increasingly powerful Linux clusters coupled with geographically
diverse end-users and increasingly large data samples typical of large
scale high energy \& nuclear physics experiments. GRID-like technology
is also making significant inroads into industrial applications and has
attracted the interest and support of well-known software manufacturers,
such as Platform Computing [2].

The Linux Farm has begun to investigate, install and support (where
appropriate) prototypes of GRID-like software that possess capabilities
that are of interest to our users, such as Ganglia, Condor and GLOBUS.

\subsection{GANGLIA}

Ganglia [3] is a full-feature, open source distributed monitoring 
software for high-performance computer clusters. It is based on a 
hierarchical design targeted at federation of clusters, and it supports 
clusters up to 2000 servers in size. Early prototypes are already 
equipped with a end-user Web interface, historical data information and 
clustering of remote systems. The monitoring data collected by ganglia 
can be used as the basis of a batch system job scheduler mechanism, 
although this has not yet been tested.  

The Linux Farm has deployed a ganglia prototype for the STAR experiment.
In the prototype, the ganglia collector has been configured to gather 
information from each of the nodes where a ganglia client daemon is 
running. This master collector then makes this information available to 
qualified external collectors. Figure 9 illustrates the ganglia deployment 
within the RCF. 

Security issues with this ganglia prototype are being investigated. 
Of particular concern, currently there is no user-friendly method to 
restrict the type and amount of information transmitted to external 
collectors. Wrap-around scripts written by RCF staff were used to 
restrict the information (see figures 7 and 8). In addition, as more 
servers are added to the ganglia master collector, scalability issues 
will become a major concern as well. The Linux Farm group plans to 
continue to test and expand the scope of the ganglia prototype where
appropriate.

\begin{figure}[h]
\begin{center}
\includegraphics[width=80mm]{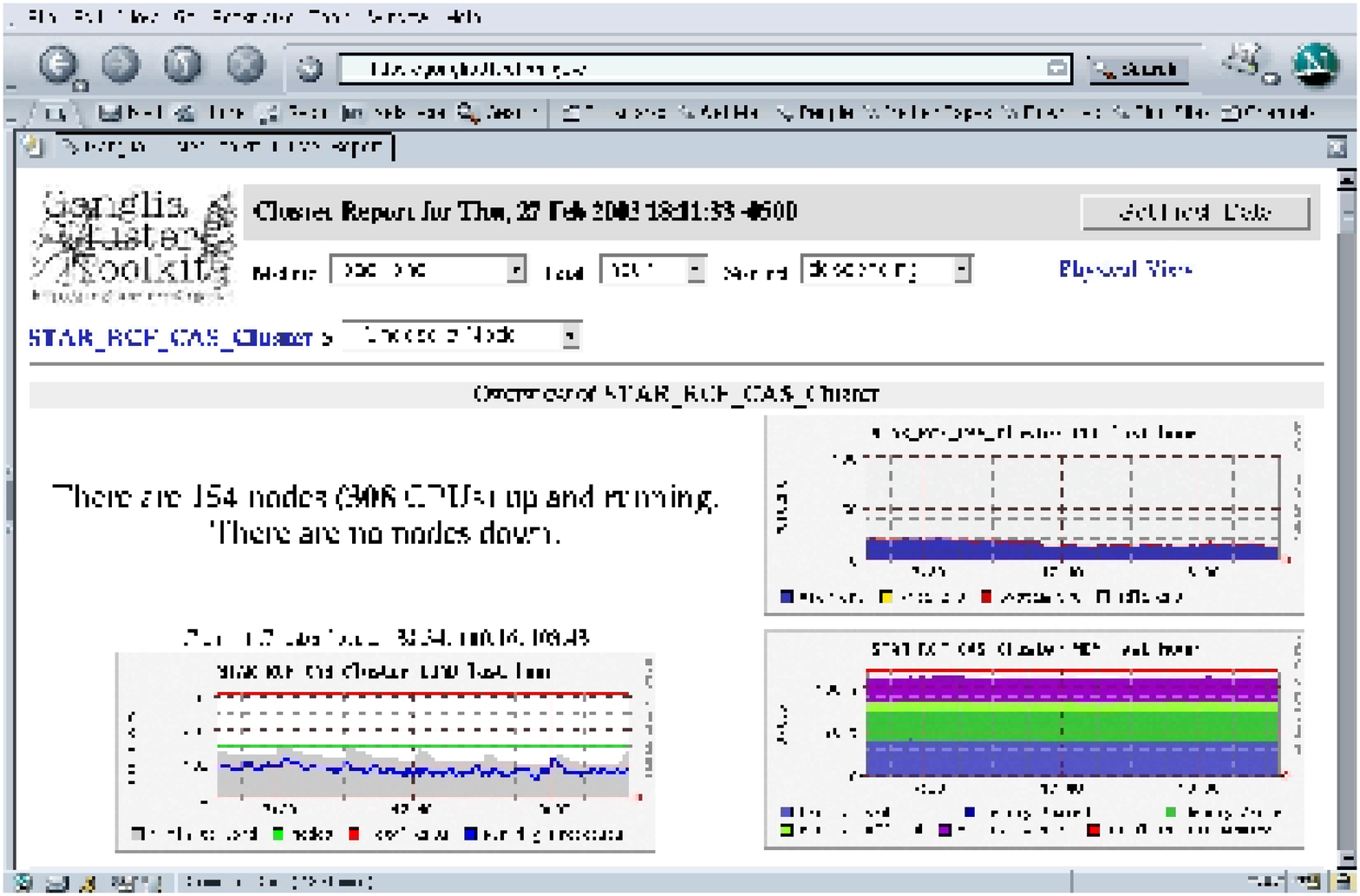}
\caption{Summary view of the ganglia prototype for STAR.}
\end{center}
\end{figure}

\begin{figure}[h]
\begin{center}
\includegraphics[width=80mm]{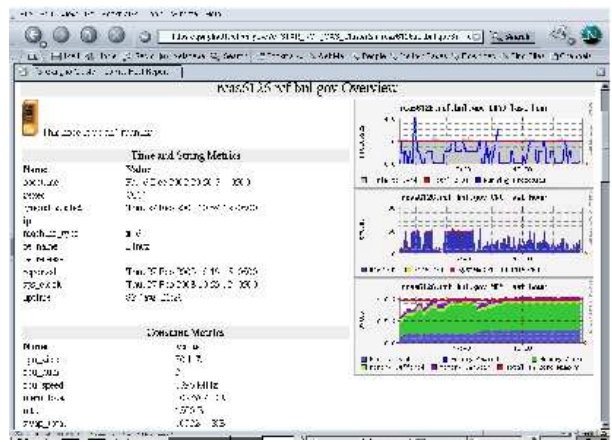}
\caption{Detailed view of a STAR node with ganglia.}
\end{center}
\end{figure}

\subsection{CONDOR}

Condor is a open-source batch software created and supported
by the University of Wisconsin [4]. Condor is a full-feature
batch software that include features such as job queuing
mechanism, configurable scheduling policy, priority scheme,
checkpoint capability and resource monitoring \& management.
Condor can be used to connect remote clusters at geographically 
diverse locations, so it is a natural fit to the GRID
computational philosophy. Condor has an interface to the GRID
via Condor-G. 

The Linux Farm group is in the midst of upgrading its 
MDS-compatible batch system to improve reliability \& scalability 
and add functionality. As part of the upgrade, Condor is being 
evaluated as a job scheduler for the new MDS-compatible batch 
system. Since media-based MDS systems is expected to play a 
considerable role at the RCF for the foreseable future, an effort
is being made to integrate Condor with the MDS-interface API 
software. The current batch system does not have an interface
to GRID-like architectures, and Condor can add this missing 
functionality via Condor-G. The basic design of the new batch 
system is shown in figure 10.

Once the prototype of the upgraded MDS-compatible batch system is
installed, Condor scalability studies will be done to understand
how performance is affected under the expected heavy usage. 

\subsection{GLOBUS \& LSF}

The ability of users to submit jobs to remote clusters has
been one of the principal motivations for the Linux Farm group
to explore interfacing our batch system with GRID-like software. 

The Linux Farm has a prototype GLOBUS [5] gatekeeper server 
that interfaces with LSF for the ATLAS experiment. Authorized 
users at remote sites can submit jobs to the gatekeeper. The 
gatekeeper interprets the GLOBUS commands and submits jobs to 
the proper LSF queues running on the ATLAS Linux cluster at the 
RCF. A diagram of the prototype is shown in figure 11. Figure 12 
shows actual LSF jobs submitted by remote users via the GLOBUS 
gatekeeper. 

Currently, the system is being expanded to include both PHENIX
and STAR experiments in RHIC, and additional GLOBUS gatekeepers 
are being brought on-line. 

\begin{figure*}[t]
\begin{center}
\includegraphics[width=160mm]{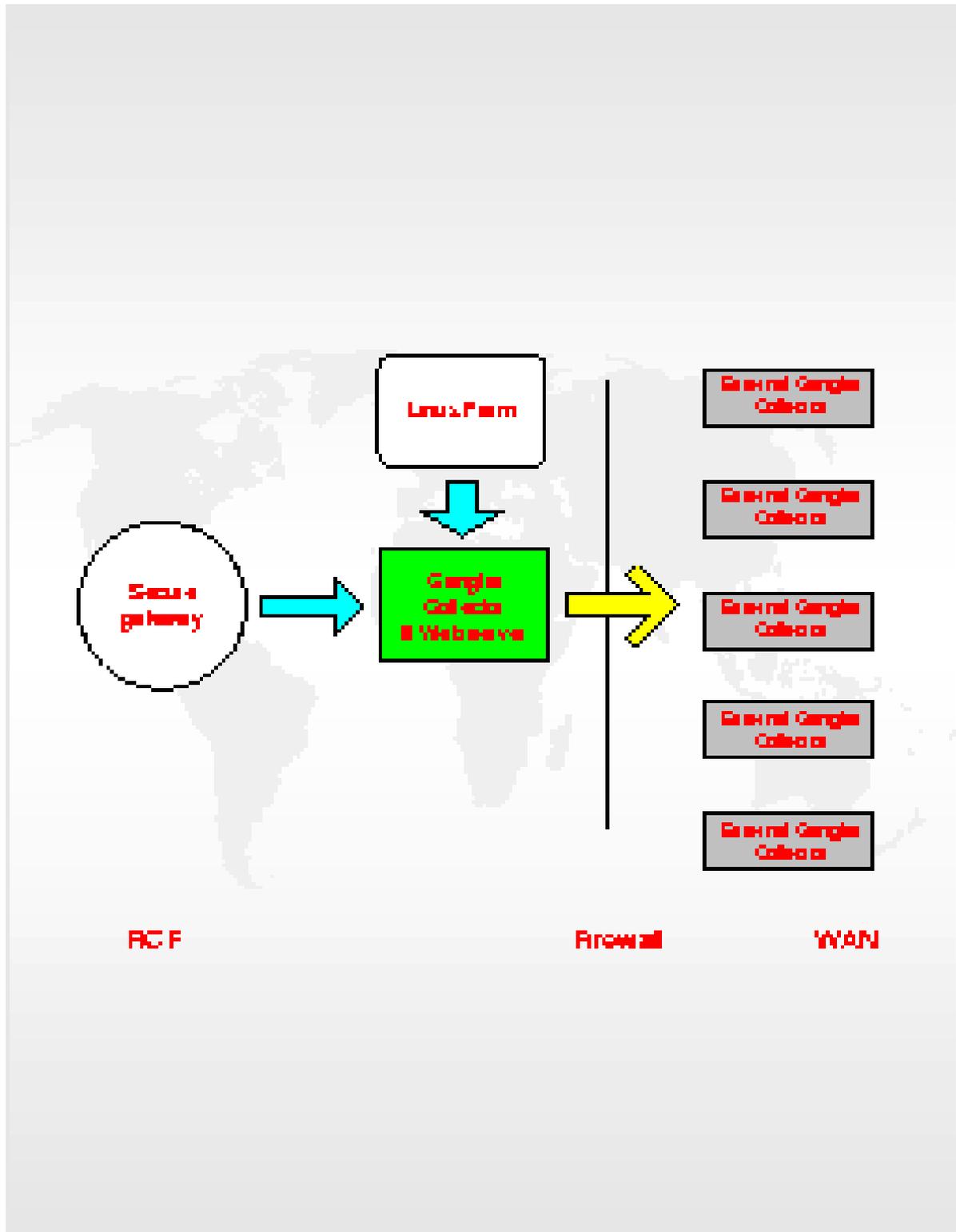}
\caption{Ganglia prototype in the RCF Linux Farm.}
\end{center}
\end{figure*}

\begin{figure*}[t]
\begin{center}
\includegraphics[width=160mm]{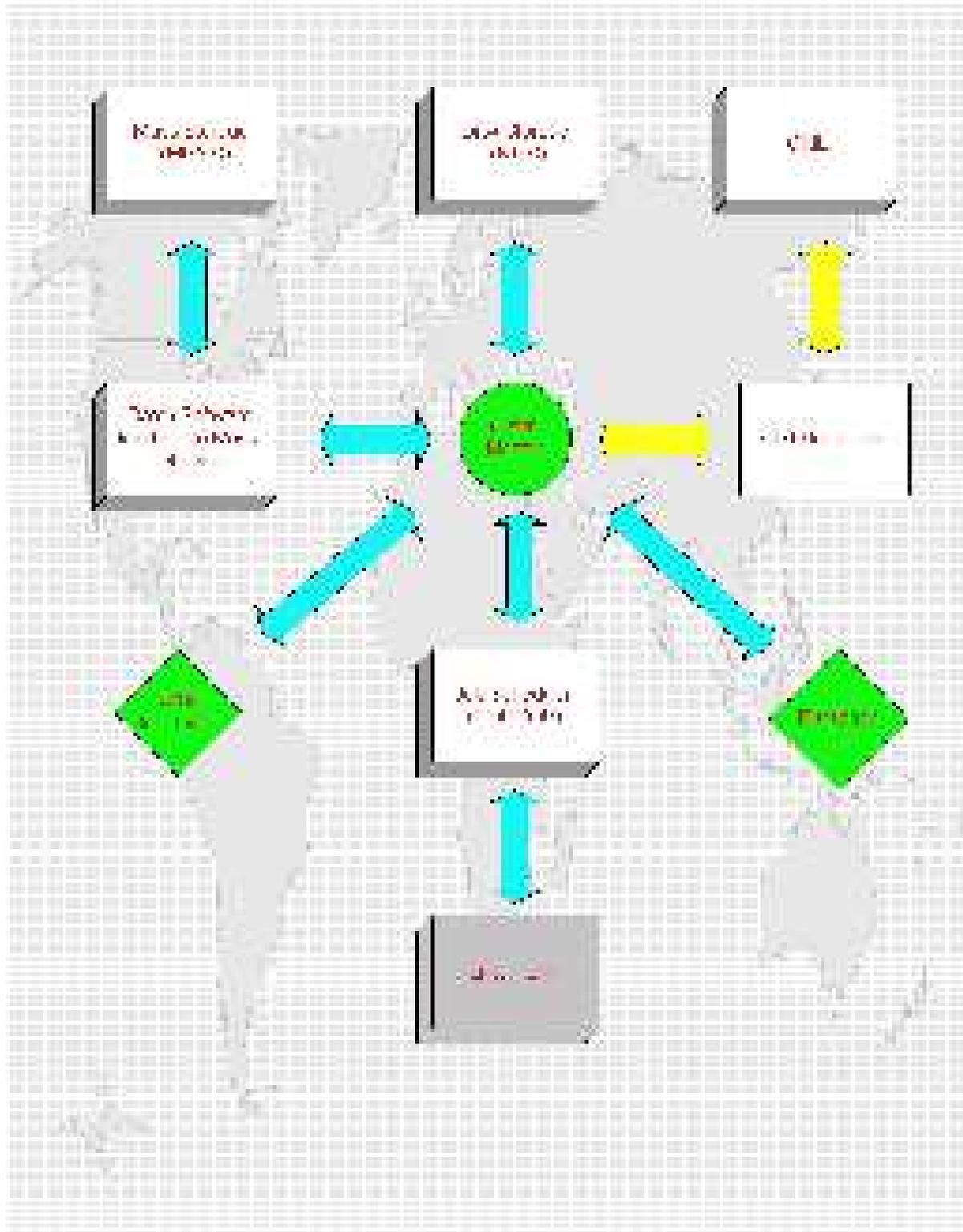}
\caption{MDS-compatible batch system at the RCF.}
\end{center}
\end{figure*}

\begin{figure*}[t]
\begin{center}
\includegraphics[width=160mm]{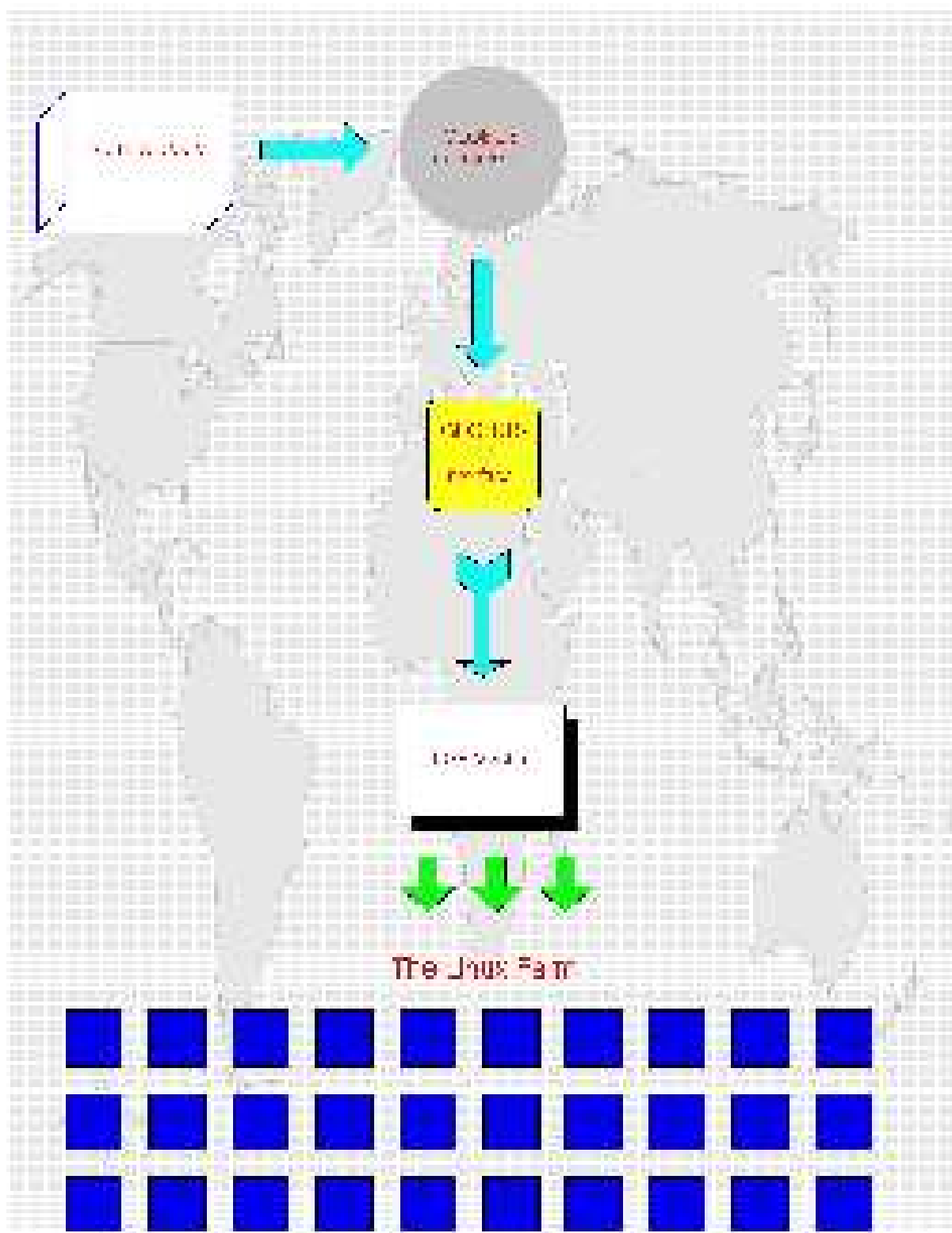}
\caption{LSF batch access via GLOBUS at the RCF.}
\end{center}
\end{figure*}

\clearpage

\begin{figure}[h]
\begin{center}
\includegraphics[width=80mm]{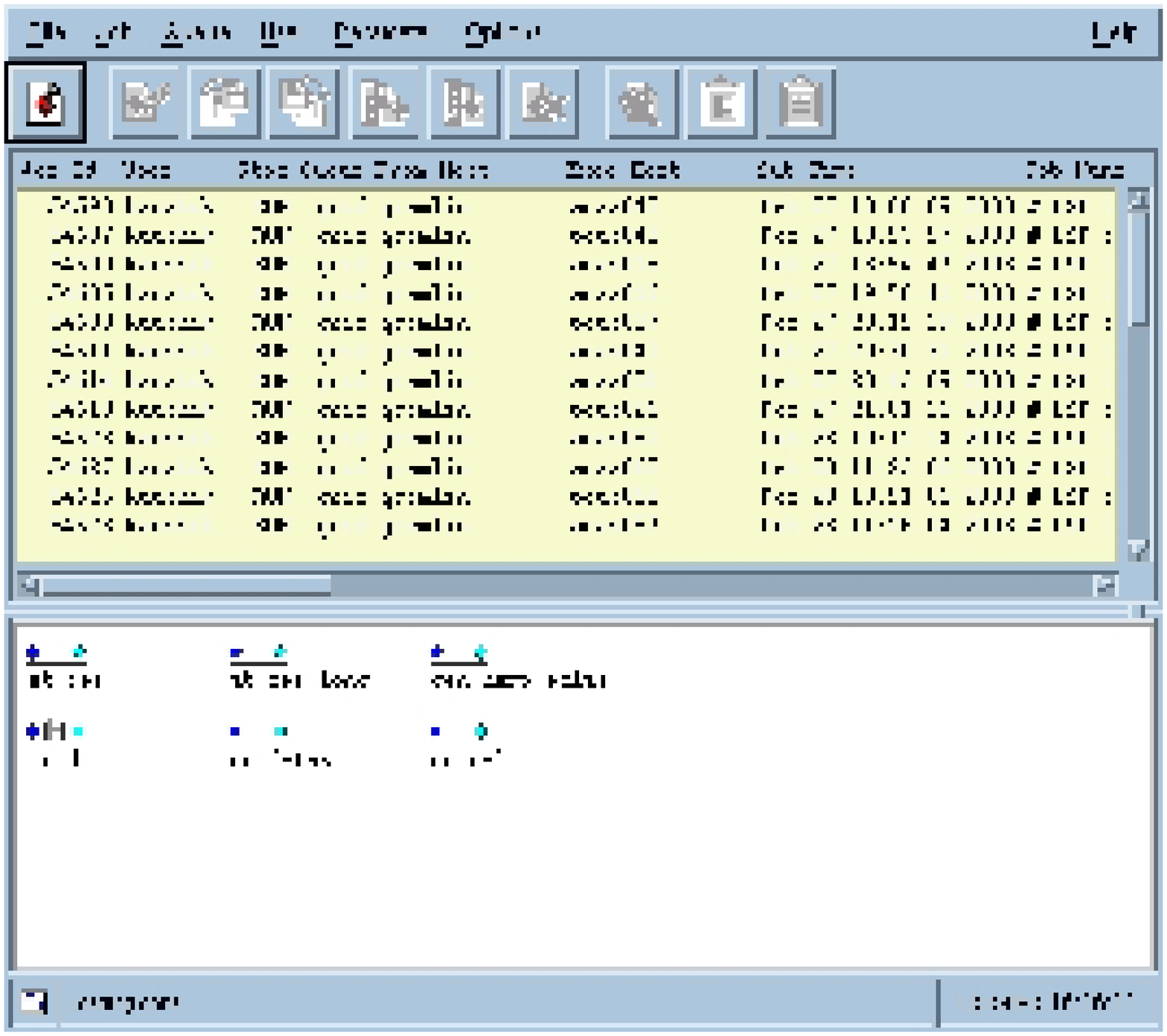}
\caption{LSF batch jobs submitted from the GLOBUS gatekeeper.}
\end{center}
\end{figure}

\section{Near-Term Plans}

In the near-term, the current prototypes are expected to expand
and slowly mature into production tools, as the outstanding issues
are resolved. 
 
Ganglia has already gone through two upgrades within the RCF, and 
the Linux Farm group is expecting that it will become part of the 
standard software packages on all its production servers in the 
near future. 

Condor has been evaluated continuously on a small number of
servers as the future job scheduler of the upgraded MDS-compatible
batch system. Many of the oustanding issues have been resolved
or are being studied, and we expect the batch system upgrade to be 
a year-long project.

The Linux Farm is currently using LSF v.4.2, and we plan to upgrade
it to LSF v.5.x together with an OS upgrade in the next few months. 
New LSF features such as advance resource reservation and GRID
membership protocols match well with the GRID computational 
architecture and can further integrate the GLOBUS gatekeepers 
with the LSF batch system.

\begin{acknowledgments}
The authors wish to thank the Information Technology Division and
the Physics Department at BNL for their support to the RCF mission.

\end{acknowledgments}


\end{document}